\documentclass[pdflatex,sn-mathphys-num]{sn-jnl}

\usepackage{graphicx}%
\usepackage{multirow}%
\usepackage{amsmath,amssymb,amsfonts}%
\usepackage{amsthm}%
\usepackage{mathrsfs}%
\usepackage[title]{appendix}%
\usepackage{xcolor}%
\usepackage{textcomp}%
\usepackage{manyfoot}%
\usepackage{booktabs}%
\usepackage{algorithm}%
\usepackage{algorithmicx}%
\usepackage{algpseudocode}%
\usepackage{listings}%

\theoremstyle{thmstyleone}%
\theoremstyle{thmstyletwo}%

\theoremstyle{thmstylethree}%

\begin{document}

\title[Article Title]{ AutoGraph: A Knowledge-Graph Framework for Modeling Interface Interaction and Automating Procedure Execution in Digital Nuclear Control Rooms}

\author[1,2]{ \sur{Xingyu Xiao}}\email{xxy23@mails.tsinghua.edu.cn}
\author[1,2]{\sur{Jiejuan Tong}}
\author[1]{ \sur{Jun Sun}}
\author[1]{ \sur{Zhe Sui}}
\author*[1]{\sur{Jingang Liang}}\email{jingang@tsinghua.edu.cn}
\author[1,2]{ \sur{Hongru Zhao}}
\author[1,2]{ \sur{Jun Zhao}}
\author[1]{\sur{Haitao Wang}}

\affil[1]{\orgdiv{Institute of Nuclear and New Energy Technology}, \orgname{Tsinghua University}, \orgaddress{ \city{Beijing}, \postcode{100084}, \country{China}}}

\affil[2]{\orgdiv{National Key Laboratory of Human Factors Engineering}, \orgaddress{ \city{Beijing}, \postcode{100094}, \country{China}}}

\abstract{
Digitalization in nuclear power plant (NPP) control rooms is reshaping how operators interact with procedures and interface elements. However, existing computer-based procedures (CBPs) often lack semantic integration with human–system interfaces (HSIs), limiting their capacity to support intelligent automation and increasing the risk of human error, particularly under dynamic or complex operating conditions. In this study, we present  AutoGraph, a knowledge-graph-based framework designed to formalize and automate procedure execution in digitalized NPP environments.AutoGraph integrates (1) a proposed HTRPM tracking module to capture operator interactions and interface element locations; (2) an Interface Element Knowledge Graph (IE-KG) encoding spatial, semantic, and structural properties of HSIs; (3) automatic mapping from textual procedures to executable interface paths; and (4) an execution engine that maps textual procedures to executable interface paths. This enables the identification of cognitively demanding multi-action steps and supports fully automated execution with minimal operator input. We validate the framework through representative control room scenarios, demonstrating significant reductions in task completion time and the potential to support real-time human reliability assessment. Further integration into dynamic HRA frameworks (e.g., COGMIF) and real-time decision support systems (e.g., DRIF) illustrates AutoGraph’s extensibility in enhancing procedural safety and cognitive performance in complex socio-technical systems.
}


\maketitle

\section{Introduction}\label{Introduction}

With the advancement of fourth-generation nuclear power technologies and the ongoing digital transformation of nuclear power plant (NPP) control rooms, modern NPPs are increasingly adopting digital instrumentation and control (I $\&$ C) systems. These developments offer improved interface richness, real-time procedural accessibility, and enhanced system observability. However, they also impose higher demands on operational safety, particularly in the context of human reliability. As the number of NPPs continues to grow worldwide, so too does the importance of ensuring robust and error-resistant operator interactions.

Despite technological progress, human error remains a leading contributor to operational incidents in nuclear facilities. According to records from the Institute of Nuclear Power Operations (INPO), approximately 48\% of reportable events in U.S. nuclear power plants can be attributed to human error \cite{zhang2025analysis}. This statistic underscores the critical need for advanced methods to prevent and mitigate human-related failures, especially in routine but safety-relevant tasks such as control room inspection rounds.

Current digital support systems in nuclear power plants, including many computer-based procedures (CBPs), often lack deep semantic understanding of human–system interfaces (HSIs). This limitation constrains their capacity to intelligently assist operators or automate procedural tasks that involve complex interface interactions—such as multi-action steps \cite{blanc2012requirements}. While knowledge graphs have been applied in system modeling and safety analysis, their dedicated use for constructing dynamic, actionable, real-time semantic models of HSIs—capable of directly supporting procedure automation within nuclear control rooms—remains an underexplored area. A critical barrier lies in the absence of machine-interpretable, formally modeled links between procedural instructions and the specific, interactive HSI elements they reference. Without these links, the execution of procedures cannot be reliably automated, especially under dynamic operating conditions where precision and contextual awareness are essential.

In addition, two major limitations persist in the current body of research. First, there is a fragmented treatment of semantics and spatial information. While semantic embedding methods are effective at capturing lexical meaning and procedural intent \cite{hall2024digitalized}, they often ignore the spatial topology of graphical user interfaces (GUIs). Conversely, qualitative spatial reasoning (QSR) techniques can model spatial relationships between interface elements, yet typically fail to account for linguistic ambiguity and procedural context. To date, an integrated semantic–spatial mapping framework that bridges this divide remains lacking. Second, there is a neglect of multi-action step complexity in both human reliability analysis (HRA) databases and GUI mapping studies \cite{mantle2019large}. In many digital procedures, a single textual instruction may implicitly compress multiple GUI operations across different screens or interface layers. However, existing methods rarely quantify the cognitive load or error likelihood introduced by such compound instructions, despite their prevalence in real-world control tasks.

To address these challenges, this paper proposes a knowledge-graph-based framework (AutoGraph) that systematically models interface structures, maps procedural tasks to interface elements, and enables automated execution of operation paths. The core idea is to construct a semantic-spatial knowledge representation of the control room interface, referred to as the interface element knowledge graph (IE-KG), that captures the names, coordinates, and parent-child relationships of interactive elements. On top of this representation, procedures are parsed and translated into sequences of interface-level actions, allowing for automatic identification of complex multi-action steps and subsequent path generation for automated execution.

The contributions of this work are fourfold:
\begin{itemize}
    \item Development of an interface tracking tool that records operator interactions with spatial and temporal resolution;
    \item Construction of a domain-specific knowledge graph to structurally represent the digital interface;
    \item A mapping mechanism that links procedural instructions to graph-based interaction paths, enabling the automatic detection of complex multi-action steps.;
    \item A prototype automation engine that executes mapped paths and demonstrates procedural automation capabilities in a simulated control room.
\end{itemize}

This work lays a foundation for scalable, interpretable, and data-integrated digital procedure support systems, and provides practical insights for enhancing operational reliability through engineering-driven HFE methodologies.

This paper is structured as follows: Section \ref{Related Work} presents the related work, Section \ref{Methodology} outlines the methodology, Section \ref{Case Study} discusses the results and evaluation of case study, and Section \ref{Conclusion} concludes with a discussion.

\section{Related Work}\label{Related Work}

\subsection{Advanced Operator Support Systems in Digital Nuclear Power Plant Control Rooms}

Modern nuclear power plant (NPP) control rooms are transitioning from analog instrumentation to digital human–system interfaces (HSIs) and computerized operator support systems to enhance efficiency, safety, and human performance \cite{boy2021human, us1981standard}. Regulatory bodies like the U.S. NRC emphasize the importance of human factors engineering (HFE) in control room design \cite{o2008human}, as exemplified by the UK EPR's integration of operational experience (OPEX) and user-centered design \cite{kirwan1996human}. However, digitalization introduces new cognitive demands on operators, especially under abnormal conditions \cite{xiao2025dynamic}, making advanced support systems increasingly critical.

While digital platforms offer powerful control capabilities—ranging from fuzzy logic to neural networks—they also add layers of complexity that can obscure system behavior and impair operator decision-making \cite{xiao2024emergency}. Traditional OPEX-driven approaches help refine system usability but may not support transformative functions like automated procedure execution. Bridging this gap calls for knowledge-driven paradigms capable of interpreting HSIs as structured, semantically rich environments.

Effective operator support must therefore move beyond static displays toward systems that can infer context, understand intent, and assist or automate procedural tasks \cite{cantucci2022collaborative}. Prior research in adaptive human–machine collaboration \cite{haesevoets2021human}—though not specific to NPPs—demonstrates the potential of cognitive architectures to tailor autonomy in real time. However, the lack of formalized, machine-interpretable task models remains a major barrier in current computer-based procedures (CBPs) \cite{xing2024enhancing}, limiting their ability to generate condition-based actions autonomously.

This motivates our proposed knowledge-graph-based approach, which aims to embed semantic understanding into HSIs and procedural logic, thereby enabling intelligent and automated support for complex control tasks in digital nuclear environments.

\subsection{Knowledge Representation in Complex Human–System Interfaces}

Ontologies and knowledge graphs (KGs) are widely adopted in safety-critical domains like nuclear power and aviation to represent complex knowledge structures, support integration, and enable reasoning \cite{meenachi2012survey, tian2022effectiveness}. Applications include risk analysis, hazard identification, and decision support through structured representations of components and safety logic. For instance, Kaur et al. \cite{wen2023systematic} emphasized knowledge reuse, while others integrated STAMP with KGs for analyzing sociotechnical safety systems \cite{simone2023extending}.

To support intelligent automation, HSIs must be semantically modeled—capturing not just visual layout but also functional roles, relationships, and procedural relevance \cite{schraagen2000cognitive}. This work contributes by encoding element coordinates, names, and hierarchies into an interface knowledge graph. Prior UI modeling efforts, such as the NRC’s use of GOMS \cite{karwowski1990framework} or the UIGuider project \cite{chen2019gallery}, focused on design usability and guideline validation. More recent efforts in robotics leverage scene graphs for spatial reasoning, offering analogs to the GUI modeling pursued here.

Despite increasing KG use in system modeling, applications enabling real-time, machine-actionable HSI semantics for procedural automation remain scarce. Most focus on post hoc analysis or UI design validation, lacking the dynamic querying and execution capability required for digital nuclear operations \cite{xing2024enhancing}. Our approach targets this gap by modeling HSIs as logic-linked, semantic graphs to support real-time automation.

A key distinction lies between world-level KGs—used for plant-level inference—and interface-level models, as developed here, which enable procedural execution by answering “how” to operate the system. These levels are complementary: the former supports functional reasoning (“why” and “with what effect”), while the latter enables concrete interface interaction, both essential for intelligent operator support in digitalized control rooms.

\subsection{Computer-Based Procedures and Automation in Nuclear Power Plant Operations}
Computer-based procedures (CBPs) were developed to replace paper-based procedures, offering benefits such as real-time plant monitoring, searchable steps, and reduced operator workload. Systems like those in the KNGR and AP1000 reactors demonstrate features like indirect component control \cite{ohara2010human}, flowchart previews, and integration with human–system interfaces (HSIs). NRC guidelines (e.g., NUREG/CR-6634) provide standards for CBP design, and the AWP (Automated Work Package) concept further advances dynamic, context-driven procedural workflows \cite{karwowski1990framework}.

Despite these advancements, CBPs still face limitations—particularly in handling complex conditional logic, integrating with HSIs, and enabling intelligent automation. Many are optimized for human readability, not machine interpretability, making it difficult to generate condition-specific prompts or execute multi-step actions \cite{xing2024enhancing}. As O’Hara et al. \cite{o2008human} and ONR assessments have noted \cite{whaley2018adapting}, the increasing complexity of digital systems raises risks of cognitive overload and procedural omission, which conventional HRA methods may underestimate.

Recent research has shifted toward intelligent procedural support capable of real-time reasoning, monitoring, and autonomous execution. Examples include agent-based automation in power systems, formal methods (e.g., Petri nets), and LLM-driven scene graph approaches in robotics \cite{zu2024language}, highlighting broader efforts to embed semantic understanding into control frameworks. However, many of these are domain-specific or lack GUI-level integration.

A key gap remains: conventional CBPs often fail to interpret the semantics of interface elements, particularly for multi-action steps. For example, a task like "normalize system X" may involve several discrete operations that standard CBPs present only as high-level text \cite{whaley2018adapting}. By contrast, our proposed method models GUI elements in a knowledge graph, enabling the system to locate relevant components, understand their roles, and execute coordinated actions. This bridges the semantic gap between procedural logic and interface operations, enhancing both automation and operator support. While AWPs represent a step forward, their potential is constrained by the limited semantic depth of current CBPs—an issue this work directly addresses through interface-encoded knowledge modeling \cite{al2018automated}.

\section{Methodology}\label{Methodology}

\subsection{Overview of Proposed Framework}

\begin{figure}[h]
\centering
\includegraphics[width=1.0 \textwidth]{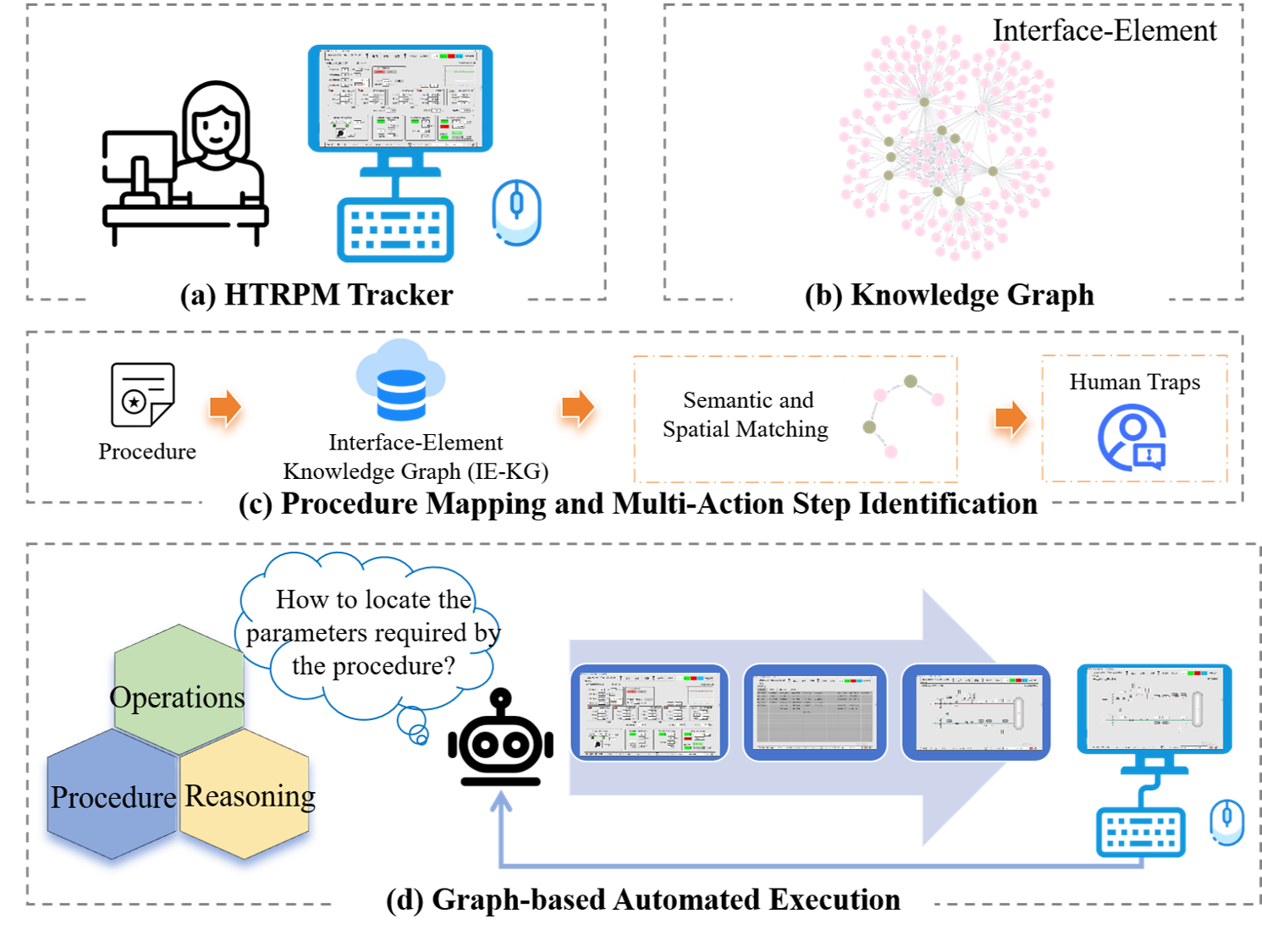}
\caption{Overview of the AutoGraph Framework}\label{workflow}
\end{figure}

To support reliable, scalable, and intelligent execution of procedures in digital nuclear control rooms, we propose an automated graph-based execution framework (AutoGraph). The workflow of AutoGraph, illustrated in Figure \ref{workflow}, comprises four core distinct phases. Specifically, Phase I involves the construction of an Interface-Element Knowledge Graph, capturing hierarchical relationships and structural attributes of the user interface. Phase II, termed procedural semantic and spatial matching, establishes semantic mappings and spatial congruences between procedural instructions and corresponding interface elements. Subsequently, Phase III focuses on dynamic human error trap detection, leveraging the semantic-spatial mappings to dynamically identify and quantify potential human error traps embedded within operational procedures.

\subsection{Phase I: HTRPM Tracker Development for Interface Element Localization} \label{Phase I}

In Phase I, we developed a customized interface tracker tailored for the HTRPM simulation environment. While optimized for this specific context, the tracker is designed as a general-purpose tool and can be readily adapted to other digital control room platforms with minimal configuration. 

The primary function of the tracker is to automatically record human-computer interaction data during simulated or real-time procedure execution. Specifically, it captures: (1) Click and keyboard press events, including left/right mouse clicks, button activations, and key inputs; (2) Screen coordinates (X, Y) corresponding to each interaction point; (3) Timestamps indicating when each action occurred; (4) Contextual interface snapshots, capturing the screen area surrounding the interaction at the moment of input. This data stream provides high-resolution temporal and spatial traces of operator behavior, enabling downstream modules to reconstruct action sequences, correlate them with procedural intent, and populate the interface element knowledge graph (IE-KG). The tool operates unobtrusively in the background and does not require intrusive modifications to the simulation software.

By ensuring consistent and structured data collection, the HTRPM tracker forms the foundation for semantic interface modeling and supports the broader goal of enabling real-time analysis, automated execution, and intelligent interface assessment in digital nuclear main control rooms.

\subsection{Phase II: Interface-Element Knowledge Graph Construction}

To enable semantic and spatial reasoning over the digital main control room interface, we construct a domain-specific interface element knowledge graph (IE-KG) that formalizes the structure, attributes, and relationships of interactive elements. The IE-KG serves as the core representation layer in our AutoGraph framework, supporting procedure-to-interface mapping, operation path inference, and multi-action step identification. The detailed workflow can be found in part (b) of Figure \ref{workflow}.

The IE-KG is defined as a directed labeled graph $G=(V, E)$, where each node $v\in V$ represents a unique interactive element in the graphical user interface, and each edge $e\in E$ denotes a hierarchical relationship between elements. Specifically, nodes are characterized by the following attributes: (1) Element Name: A human-readable label extracted from the simulation interface or metadata tags (e.g., "Nuclear Island System", "Electrical System"); (2) Coordinates: Two-dimensional screen position (x,y) that reflects the spatial location of the element. Edges in the graph are used to represent the hierarchical relations, denoting interface nesting (e.g., “Conventional Island System” → “0PCBDW001”).

The IE-KG is manually constructed based on the tracker data described in Section \ref{Phase I} and HTRPM simulator interface. As the tracker captures element positions, names, these data points are parsed and inserted into the graph database. The following two steps outline the process: (1) element extraction: interactive elements are identified and instantiated as nodes, with attributes populated from tracker outputs. (2) hierarchy resolution: parent-child structures are resolved based on window/frame identifiers and GUI layout information.

This graph-based representation enables efficient querying and reasoning over interface components, such as identifying all elements involved in a specific procedural task, detecting ambiguous or confusing interface clusters, or computing the shortest actionable path across multiple views. Furthermore, the graph can be incrementally updated as new interface elements are encountered or reconfigured during system upgrades, ensuring long-term adaptability. The IE-KG thus forms the semantic backbone of the proposed automation system, bridging the gap between procedural text and the dynamic visual interface of digital NPP main control rooms.

\subsection{Phase III: Procedure Mapping and Multi-Action Step Identification} \label{Phase III}

In digitalized control environments, operating procedures are often structured as textual or tabular instructions that must be interpreted and executed through interactions with graphical user interfaces (GUIs). However, there exists a semantic gap between the procedural representation (written in human-readable language) and the interface structure (represented by screen elements with spatial and hierarchical logic). To address this challenge, we developed a mapping mechanism that translates procedural steps into executable paths within the interface element knowledge graph (IE-KG), enabling automated execution and complexity analysis.

The detailed workflow is illustrated in part (c) of Figure~\ref{workflow}. Given an input procedure, the system first searches within the constructed IE-KG to identify the corresponding graph-based path. If the identified path involves multiple intermediate nodes, it is classified as a multi-action step. This indicates that the user must perform several sequential interactions to reach the target interface element, potentially including unnecessary operations. Such multi-action steps can impose additional cognitive load and increase the likelihood of human error. These could be used to identify high-risk procedural segments, prioritize automation, and support human reliability evaluation.

By bridging procedural semantics with spatial interface representations, this mapping module enables robust execution planning and lays the groundwork for interface-aware risk assessment in digitalized nuclear operations.

\subsection{Phase IV:  Automated Execution Module} \label{Phase IV}

Building upon the procedure-to-interface mapping outlined in Section \ref{Phase III}, we implement an automated execution module designed to carry out procedural steps in a precise, repeatable, and interface-aware manner. This module acts as the final operational layer of the proposed AutoGraph framework, transforming the inferred action paths into real-time control commands within the digital control room environment.

Figure~\ref{workflow} (d) presents the architecture of the automated execution module and its integration with the constructed IE-KG. Upon receiving a procedural instruction, the system first identifies the corresponding graph node path within the IE-KG that matches the textual description of the procedure. Subsequently, the system automatically triggers click operations on the digital interface by controlling the underlying computer system, thereby executing the required actions in a manner consistent with human interaction.

This automated execution module transforms static procedures into executable, interface-aware action sequences, thereby bridging semantic, spatial, and temporal gaps between procedural intent and system operation. It forms a critical pillar of the proposed framework’s ability to support intelligent, scalable, and reliable automation in complex human-machine environments.



\section{System Implementation and Demonstration}\label{Case Study}

A HTR-PM full-scope simulator (comprising two reactors with a combined capacity of 210 MWe) was employed in this study. A cohort of six graduate students, including two masters and four doctoral candidates, participated in experiments to evaluate the AutoGraph framework.

\subsection{Case Description and Experimental Setup}\label{Case Description and Experimental Setup}

The experimental procedures comprised five task scenarios, as detailed in Table \ref{procedures}. Each scenario was subdivided into 7, 7, 6, 5, and 4 procedural sub-steps, respectively. Participants were instructed to follow these sub-steps sequentially and perform corresponding operations on the HTGRSim simulation platform.

The study recruited six graduate students majoring in nuclear science and technology, including two master's and four doctoral candidates. Prior to the experiment, the principal investigator provided comprehensive training and ensured that all participants became familiar with the HTGRSim simulator. To encourage authentic error-inducing behavior, a time-based incentive mechanism was introduced: a performance-based reward was offered to the participant who completed all tasks in the shortest time. This design intentionally introduced time pressure and psychological stress to simulate the cognitive and operational demands encountered by operators during real-world emergency scenarios. Each participant spent approximately 20 to 40 minutes completing the tasks, resulting in a total data collection time of around 3 hours.

Three categories of data were collected: (1) full-screen recordings of the simulator interface during task execution; (2) real-time cursor movement and mouse-click coordinates, captured using a custom-developed tool (tracker.exe); and (3) external video footage recorded by an additional camera system to document the complete experimental process from an observational perspective.

To better reflect realistic operational practices, we incorporated a human performance tool (HPT) commonly used in nuclear power plants, namely, the two-step verification card. Participants were instructed to adopt a dual-verification approach: during the procedure reading phase, they circled each intended action, and after execution, they marked an “X” within the corresponding circle. This human performance tool is designed to mitigate the risk of execution of omission (EOO), a common type of human error. In our experiment, the implementation of this tool proved effective, and, as expected, no EOO instances were observed throughout the experimental sessions.

\begin{table}[h]
\caption{Procedures Used for Experimental Validation of the AutoGraph Framework}\label{procedures}%

\begin{tabular}{p{0.02\linewidth} p{1.15 \linewidth} }

\toprule
\textbf{No}  & \textbf{Content} \\
\midrule
\textbf{1} & 
1. Check whether the value of parameter 2LBA10CP801C under Nuclear Island System 2 LAB DW001 is 13.86 MPa.

2. Check whether the value of parameter 2LAB10CF801D under Nuclear Island System 2 LAB DW001 is 88.49 kg/s.

3. Check whether the value of parameter 0KBE10CP007 under Auxiliary System 0 KBE DW101 is 7018684.0 Pa.

4. Check whether the value of parameter Outlet Pressure of Open-Cycle Cooling Water Pump under Conventional Island System 0 PCB DW001 is 0.337 MPa.

5. Check whether the value of parameter Outlet Temperature of No.1 Steam Generator under Conventional Island System 0 LBH DW001 is 534.6°C.

6. Check whether the value of parameter 1JET01CP001 under Nuclear Island System 1 JET DW001 is 0.101 MPa.

7. Check whether the value of parameter Power Factor under Electrical System 0 ELE DW002 is 0.767.
\\
\midrule
\textbf{2} & 
1. Check whether the value of parameter 2LBA10CP801A under Nuclear Island System 2 LAB DW001 is 13.86 MPa.

2. Check whether the value of parameter 1JET01CT001 under Nuclear Island System 1 JET DW001 is 20.0°C.

3. Check whether the value of parameter 2LAB10CF001 under Conventional Island System 0 LBH DW001 is 88.49 kg/s.

4. Check whether the value of parameter Cooling Water Outlet Temperature of Generator Air Cooler under Conventional Island System 0 PCB DW001 is 20.0°C.

5. Check whether the value of parameter 0KBE10CT001 under Auxiliary System 0 KBE DW101 is 257.3°C.

6. Check whether the value of parameter 0KBE10CP001 under Auxiliary System 0 KBE DW101 is 9.0 kPa.

7. Check whether the value of parameter Terminal Voltage under Electrical System 0 ELE DW002 is 21.0 kV.
\\
\midrule
\textbf{3} & 
1. Check whether the value of parameter 1JET01CL001 under Nuclear Island System 1 JET DW001 is 6.0 m.

2. Check whether the value of parameter 2LAB10CF801B under Nuclear Island System 2 LAB DW001 is 88.49 kg/s.

3. Check whether the value of parameter 0KBE10CT002 under Auxiliary System 0 KBE DW101 is 283.9°C.

4. Check whether the value of parameter Excitation Current under Electrical System 0 ELE DW002 is 5366.7 A.

5. Check whether the value of parameter 1LBA10CP801A under Conventional Island System 0 LBH DW001 is 13.96 MPa.

6. Check whether the value of parameter Phase A Current of Open-Cycle Cooling Water Pump under Conventional Island System 0 PCB DW001 is 18.4 A.
\\
\midrule

\textbf{4} & 
1. Check whether the value of parameter 2LB10AA404 under Nuclear Island System 2 LAB DW001 is 0.0 kg/s.

2. Check whether the value of parameter 0KBE10CP005 under Auxiliary System 0 KBE DW101 is 30.0 kPa.

3. Check whether the value of parameter 2LBA10CP701A under Conventional Island System 0 LBH DW001 is 15.96 MPa.

4. Check whether the value of parameter Cooling Water Outlet Temperature of Vacuum Pump Cooler under Conventional Island System 0 PCB DW001 is 15.1°C.

5. Check whether the value of parameter Excitation Voltage under Electrical System 0 ELE DW002 is 548.4 V.
\\
\midrule
\textbf{5} & 
1. Check whether the value of parameter 2LBA10CP801B under Nuclear Island System 2 LAB DW001 is 13.86 MPa.

2. Check whether the value of parameter 0KBE10CP004 under Auxiliary System 0 KBE DW101 is 0.0 MPa.

3. Check whether the value of parameter Bypass Steam Temperature No.1 under Conventional Island System 0 LBH DW001 is 80.0°C.

4. Check whether the value of parameter Generator Reactive Power under Electrical System 0 ELE DW002 is 158.7 MVar.
\\
\botrule
\end{tabular}
\end{table}

\begin{figure}[h]
\centering
\includegraphics[width=0.6\textwidth]{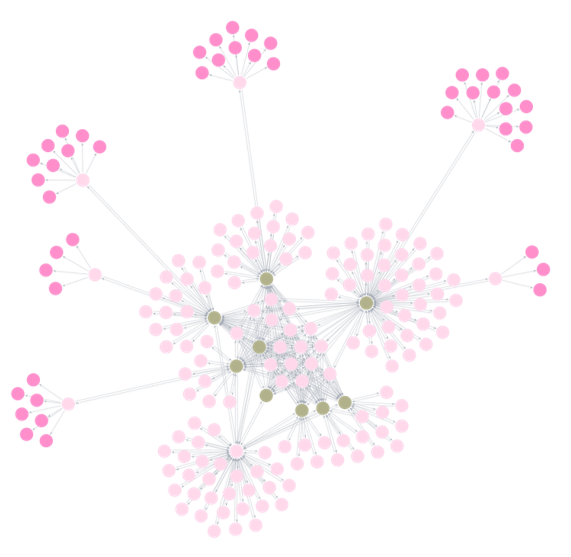}
\caption{Interface-Element Knowledge Graph (IE-KG) Derived from the HTGRSim Interface}\label{KG}
\end{figure}

We constructed a partial interface-element knowledge graph (IE-KG) for selected panels of the HTGRPMSim. The resulting knowledge graph is illustrated in Figure \ref{KG}, where different colors indicate hierarchical levels of interface accessibility. Dark green nodes represent the top-level elements that are directly visible and interactable without any user navigation. Light pink nodes correspond to intermediate layers, while dark pink nodes denote the deepest level elements, which require multiple user interactions (e.g., sequential clicks) to access.

\subsection{Dynamic Detection of Multi-Action Human Error Traps}

For the identification of multi-action procedure, it is sufficient to locate the corresponding interface elements mapped from the target procedural instruction and extract the associated navigation path. If the mapped path contains multiple sequential actions, the step can be unequivocally classified as a multi-action step. This classification is highly straightforward and does not typically require additional validation, with detection accuracy reaching 100\%.

As an illustrative example, consider the procedural task: “Check whether the parameter 2LBA10CP801C under Nuclear Island System 2 LAB DW001 equals 13.86 MPa.” The extracted navigation path is shown in Figure \ref{path_combine}. The interface traversal begins at the initial interface, proceeds to the ‘Flowchart’ module, then navigates to ‘Nuclear Island System’, continues to ‘2LABDW001’, and finally reaches the target element ‘2LBA10CP801C’. Each step in the path corresponds to an interface node with associated semantic labels and spatial coordinates, thereby validating the multi-action structure of the task.

\begin{figure}[h]
\centering
\includegraphics[width=1.0\textwidth]{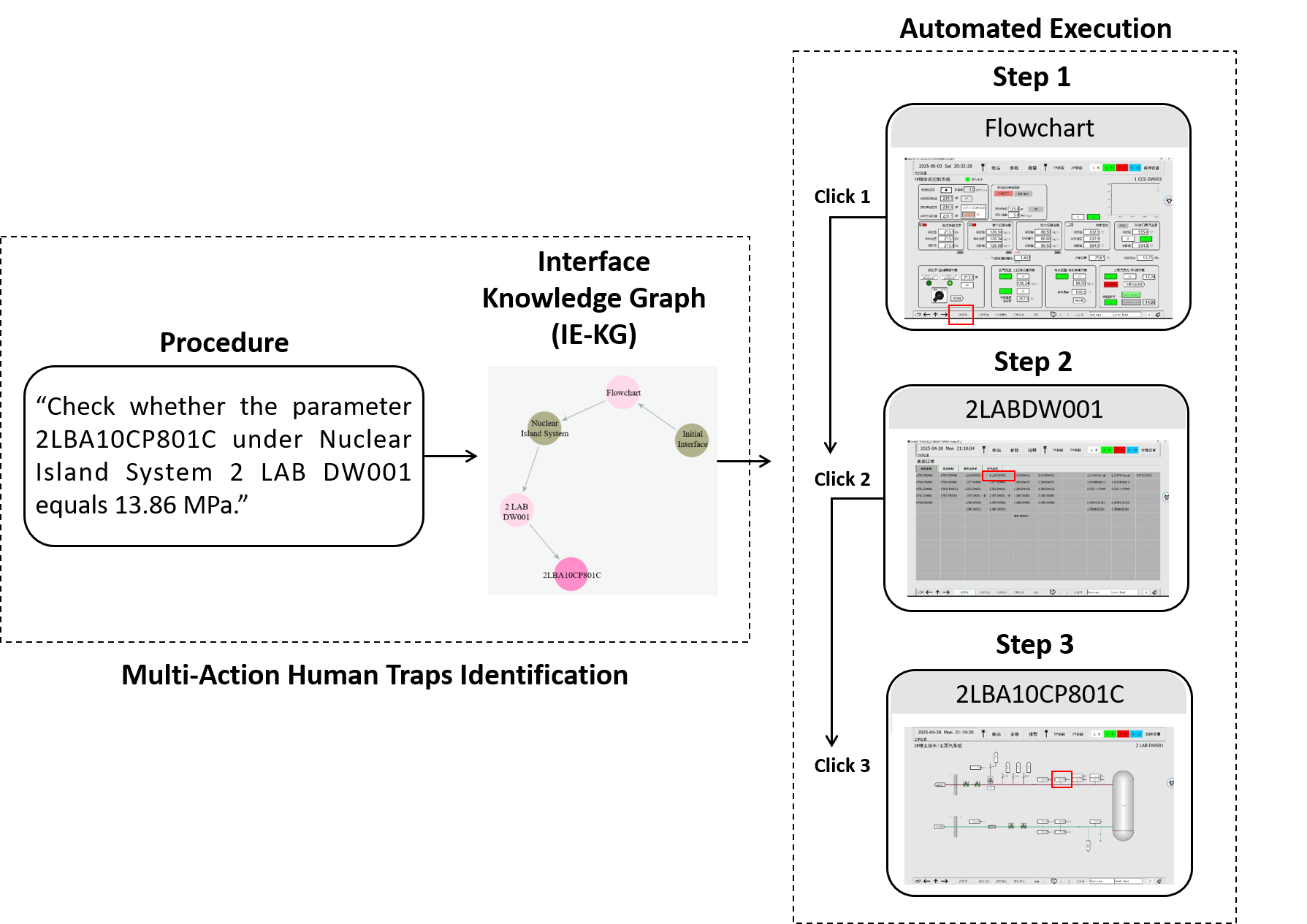}
\caption{Interface navigation path corresponding to the procedural task: “Check whether the parameter 2LBA10CP801C under Nuclear Island System 2 LAB DW001 equals 13.86 MPa.”}\label{path_combine}
\end{figure}

The identified path corresponds directly to the operator’s actual interaction sequence, aligning with both the collected interface data and real-world operational behavior. This consistency demonstrates that, when combined with the constructed interface-element knowledge graph (IE-KG), our model is capable of accurately recognizing procedural steps involving multiple actions.

\subsection{Automated Execution for Path-Based Parameter Search} \label{Automated Execution for Path-Based Parameter Search}

Building on the interface-action paths identified in the previous phases, the automated execution module developed in Phase IV is leveraged to streamline and automate the process of parameter search within digital control room environments. In conventional settings, such tasks often require operators to manually navigate through multiple interface layers, locate the correct control panels, and iteratively adjust or observe system parameters. This process can be time-consuming, error-prone, and cognitively demanding.

By utilizing the pre-mapped operation paths generated through the interface element knowledge graph (IE-KG), the execution engine can reproduce the required navigation and interaction sequences without manual intervention. Specifically, the module programmatically simulates a sequence of mouse clicks, keyboard presses, or touch interactions that correspond to the original operator procedure for locating and retrieving specific parameter values. This capability enables the system to: (1) automatically access deeply nested interface elements relevant to target parameters; (2) execute complex multi-action sequences that would otherwise require expert operator knowledge; (3) collect and store parameter values for further analysis, visualization, or system feedback.

A representative demo scenarios illustrating this capability are presented in Figure~\ref{path_combine}. These examples demonstrate the system’s ability to complete end-to-end parameter search tasks across different panels and interface states, including tab navigation, dynamic field recognition, and condition-based value confirmation. The implementation of path-based automation for parameter search not only reduces operational workload but also provides a scalable mechanism for batch data extraction, system validation, and integration with real-time decision support tools. Furthermore, automation in this context reduces the likelihood of human error. Lastly, rather than accessing parameter values programmatically, this execution approach simulates human-like interaction through automated clicks and interface navigation. Such a process mirrors the operator's actual workflow, enabling the observation of system transitions and contributing to higher credibility and operator trust.

\section{Discussions}
\subsection{Comparison of AutoGraph and Human Performance in Procedure Execution}

\begin{figure}[h]
\centering
\includegraphics[width=1.0\textwidth]{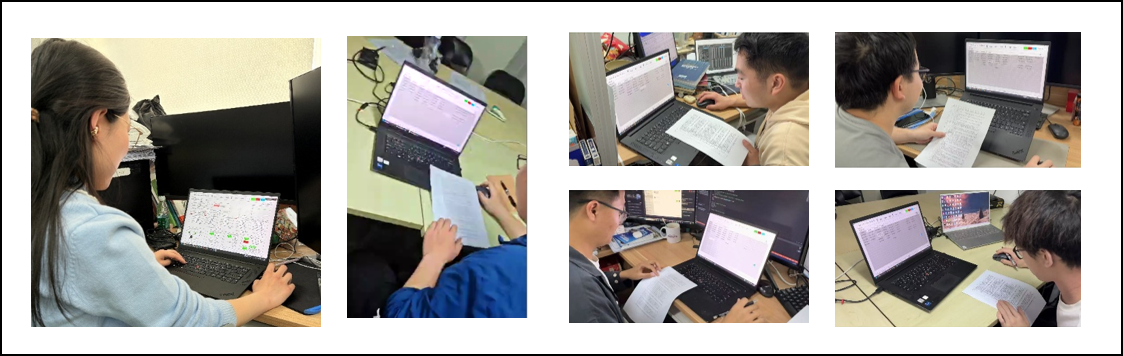}
\caption{Human-in-the-Loop Experiment Conducted in a Digital Control Room Simulation}\label{exp}
\end{figure}

As described in Section~\ref{Case Description and Experimental Setup}, we conducted a series of experiments involving real human participants performing operational tasks in a simulated digital control room environment (Figure \ref{exp}). These experiments were designed to capture detailed timing data for each step of the procedure execution, enabling direct comparison with the automated execution module.

Each experiment lasted approximately 20 to 40 minutes, depending on the complexity of the scenario and the participant’s familiarity with the interface. A noteworthy challenge in this phase was the significant post-processing effort required to extract, synchronize, and integrate the experimental data. Specifically, aligning the data collected by the interface tracker with the screen recordings and supplementary video footage on a common time axis was a labor-intensive process. On average, processing a single session required nearly eight hours of manual work, highlighting the considerable overhead involved in human data interpretation and integration.

\begin{figure}[h]
\centering
\includegraphics[width=1.0\textwidth]{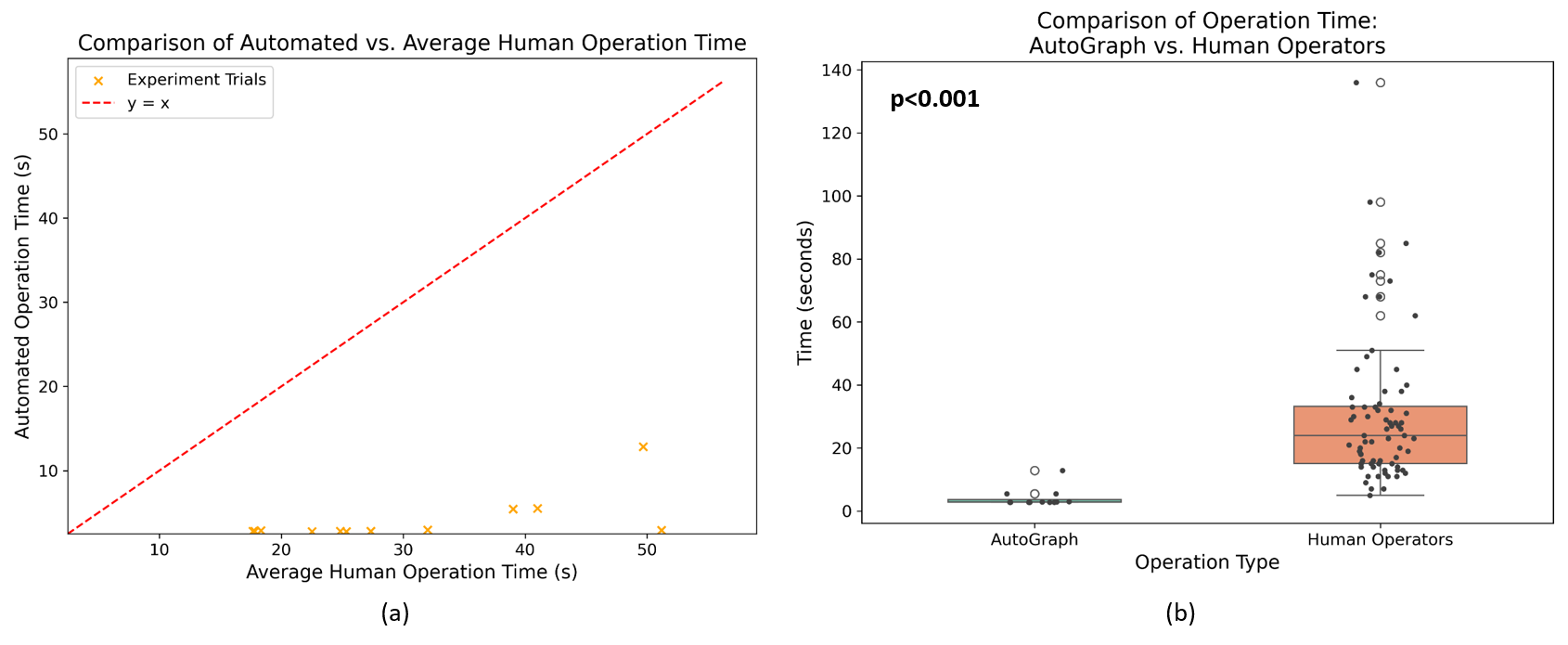}
\caption{Comparison between AutoGraph and human task completion time.}\label{compare_time}
\end{figure}

To evaluate the efficiency of the automated framework AutoGraph, we compared the task completion times between the automated system and human participants. As shown in Subfigure~\ref{compare_time} (a), the dashed line represents the equality reference $(y = x)$. All data points fall below this line, indicating that the automated approach consistently outperformed human operators in terms of time efficiency across all tested scenarios.

For statistical analysis, we applied the Mann–Whitney U test \cite{mcknight2010mann}, a non-parametric method suitable for comparing two independent samples with potentially non-normal distributions. The goal was to assess whether the observed difference in task completion times between the Automated Software and Human Operators is statistically significant or likely due to random variation. The results, presented in Subfigure~\ref{compare_time} (b), show a p-value < 0.001, providing strong statistical evidence that the automated system significantly reduces operation time compared to manual human execution.

In summary, the experimental findings demonstrate not only the operational feasibility of the proposed automation framework but also its substantial advantage in execution efficiency, supporting its value in practical deployment for complex procedural tasks in digitalized nuclear control environments.

\subsection{Integration with existing DHRA tools}\label{Integration with existing DHRA tools}

Many recent studies have focused on dynamic human reliability analysis (dynamic HRA), including tools such as human unimodel for nuclear technology to enhance reliability (HUNTER) \cite{boring2022hunter}, cognitive-mechanistic human reliability analysis framework (COGMIF) \cite{xiao2025cognitive}, goals operators methods and selection rules (GOMS) \cite{setthawong2019updated}, and queueing network-model human processor (QN-MHP) \cite{chen2024influence}. These approaches aim to achieve dynamic HRA by modeling human performance at a finer granularity and simulating operator behavior in real time. However, a common characteristic of these tools is the need to decompose broad and abstract procedural tasks into detailed, step-by-step operator actions.

This decomposition process poses significant challenges due to the sheer volume and complexity of procedures in nuclear power operations. In this context, our proposed AutoGraph framework offers a promising solution: by leveraging the constructed interface element knowledge graph (IE-KG), AutoGraph enables the automatic refinement of broad tasks into concrete operational steps based on interface-level information.

To demonstrate the integration of AutoGraph into existing dynamic HRA tools, we take COGMIF as an example. Proposed by Xiao et al. \cite{xiao2025cognitive} in 2025, COGMIF is essentially a hybrid model that combines the ACT-R cognitive architecture to automate the time estimation process of the IDHEAS-ECA methodology. Therefore, AutoGraph is first used to automatically transform high-level procedural tasks into detailed interface-linked operations. These refined steps are then used to build ACT-R-based cognitive models for estimating available operation time. Finally, the outputs are incorporated into the IDHEAS-ECA framework to support a more dynamic and context-sensitive human reliability analysis.

\begin{figure}[h]
\centering
\includegraphics[width=0.32\textwidth]{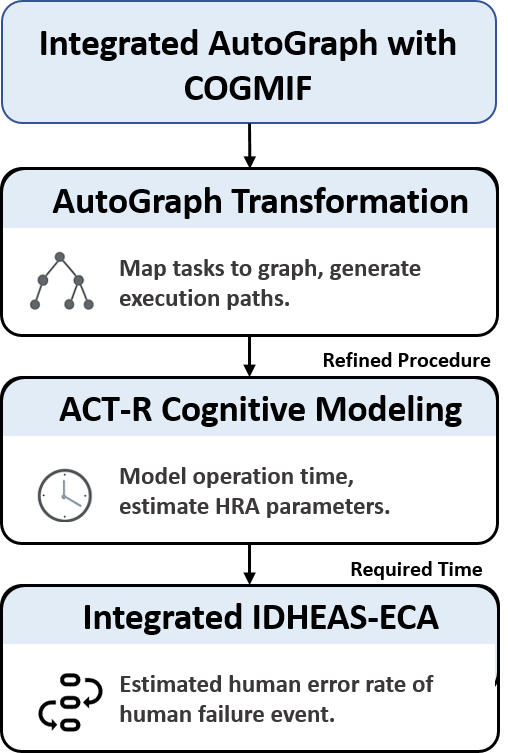}
\caption{Overview of the AutoGraph-based framework for transforming procedures into executable operator-interaction paths, and integration into the dynamic HRA system.}\label{path_combine}
\end{figure}

\begin{figure}[h]
\centering
\includegraphics[width=1.0 \textwidth]{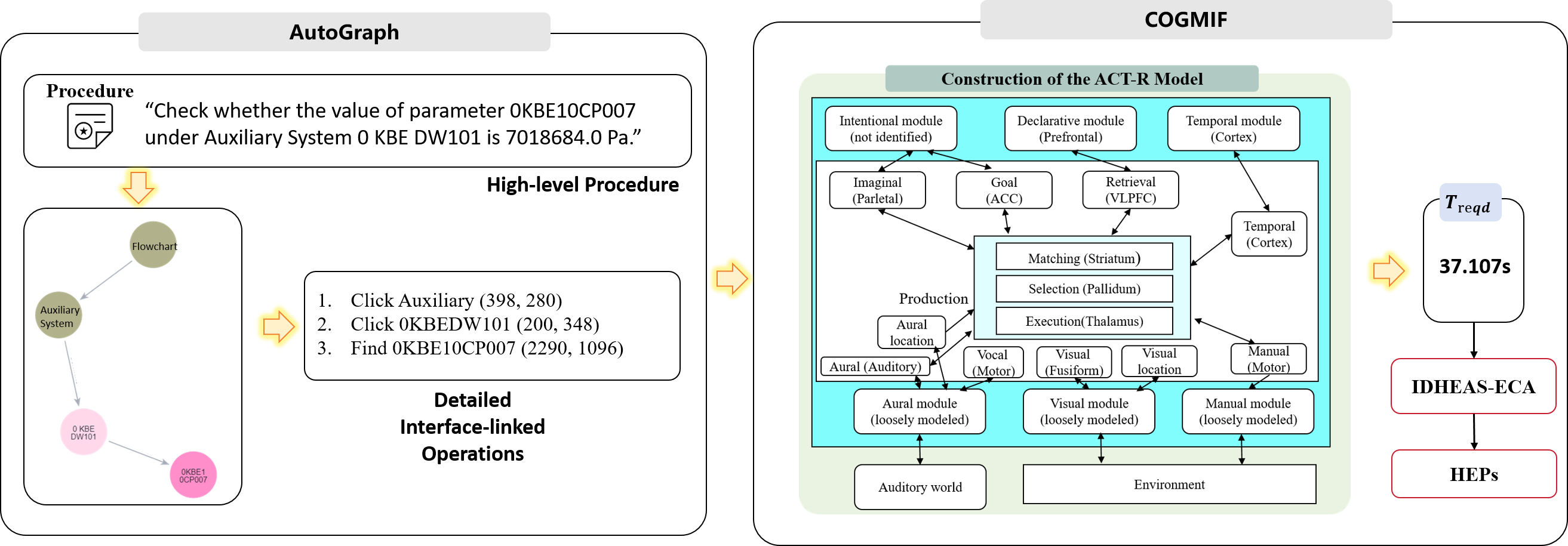}
\caption{Integration of AutoGraph into the COGMIF Framework.}\label{DHRA2}
\end{figure}

During the experimental process, we observed two instances of human error associated with the task "Check whether the parameter value of 0KBE10CP007 in Nuclear Island Auxiliary 0 KBE DW101 is 7018684.0 Pa." Accordingly, we selected this procedure as a representative case to demonstrate the application of our proposed framework. As shown in Figure~\ref{DHRA2}, the procedural steps were first mapped using the AutoGraph framework. Based on the mapped path, we constructed a cognitive model using ACT-R, incorporating Fitts' Law to account for motor movement and interface interaction. The model predicted the required task completion time ($T_\text{reqd}$) to be 37.107 seconds. Subsequently, we applied the IDHEAS-ECA methodology to compute the human error probability (HEP), which was estimated to be $8.2 \times 10^{-3}$.

\subsection{Integration with existing real-time risk-informed decision support systems} \label{Integration with existing real-time risk-informed decision support systems}

At present, a variety of decision support systems have been developed, such as the real-time online decision support system for nuclear emergencies (RADOS) and others \cite{xiao2024emergency}. However, many of these systems are large-scale and complex in structure. In recent years, fault diagnosis \cite{xiao2023enhancing,qi2024multimodal} has emerged as a core component within real-time decision support, particularly in safety-critical domains like nuclear power. Numerous studies have focused on identifying initial events by analyzing plant sensor data to support early-stage decision-making.

In this subsection, we take the fault diagnosis module as a representative example to illustrate the integration of our AutoGraph framework with existing real-time, risk-informed decision support systems.

\begin{figure}[h]
\centering
\includegraphics[width=0.9 \textwidth]{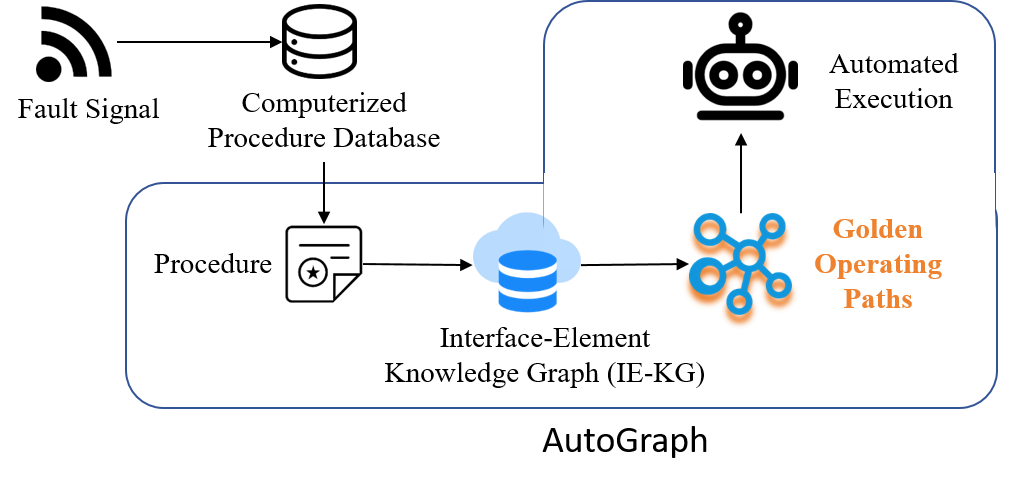}
\caption{Overview of the AutoGraph-based framework for transforming procedures into executable operator-interaction paths, and integration into the dynamic HRA system.}\label{DSS}
\end{figure}

The overall workflow is depicted in Figure~\ref{DSS}. Once a nuclear power plant operator identifies the fault type and decides to initiate the corresponding procedure, the relevant procedure is retrieved from the computerized procedure database based on diagnostic signals. The target operational paths are then derived using the Interface Element Knowledge Graph (IE-KG) developed in this study. These paths serve as a reference standard for evaluating whether the actual operator actions conform to procedural expectations, thereby enabling the systematic collection of human reliability data.

\begin{figure}[h]
\centering
\includegraphics[width=1.0 \textwidth]{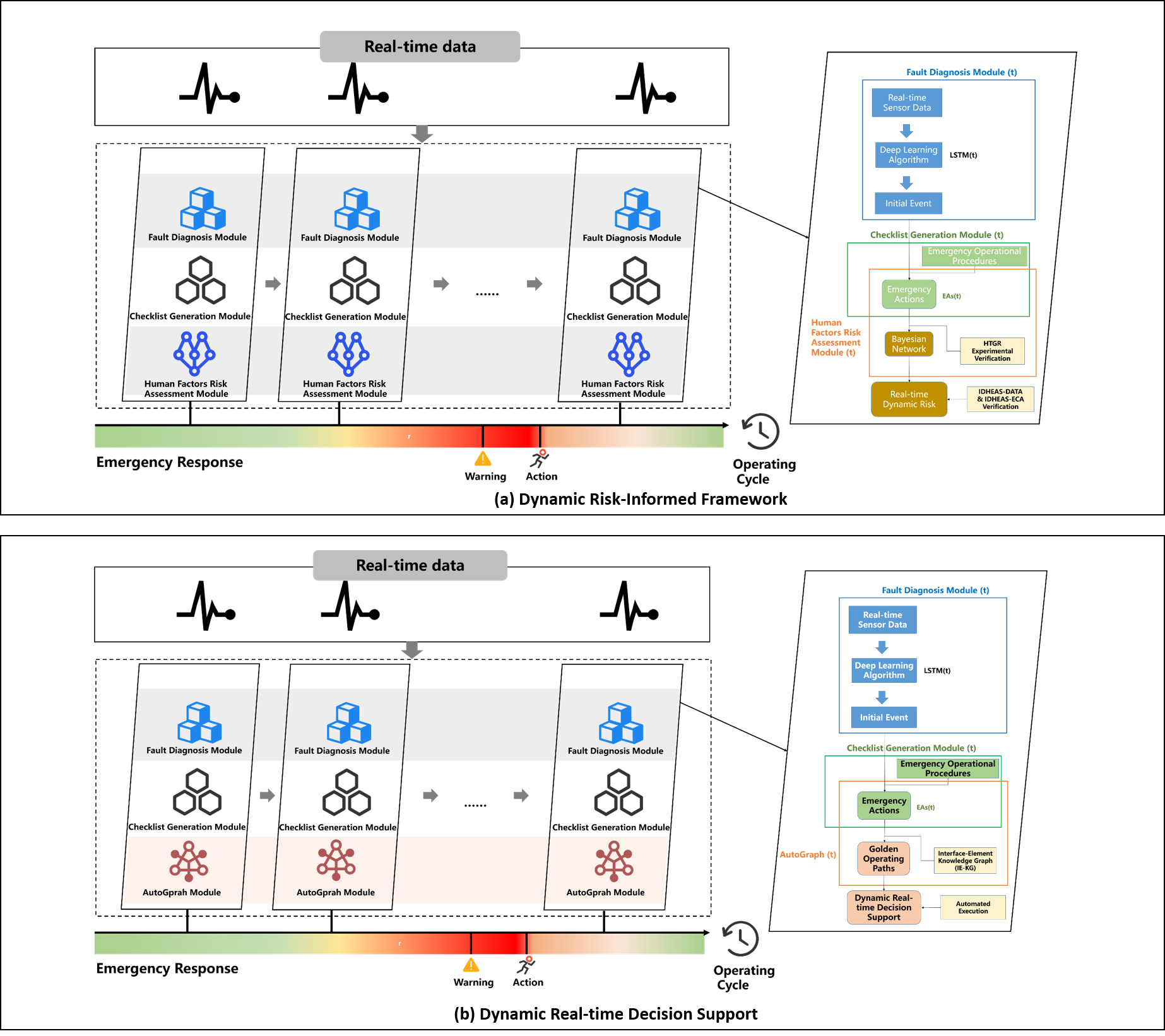}
\caption{Integration of AutoGraph into the DRIF Framework.}\label{DRIF}
\end{figure}

To demonstrate the applicability of our proposed framework, we reference the work of Xiao et al. \cite{xiao2025dynamic}, which introduced the DRIF (dynamic risk-informed framework) for human reliability analysis. As shown in Figure \ref{DRIF} (a), this framework consists of three integrated components: the fault diagnosis module, the checklist generation module, and the human factors risk assessment module. In their case study, the authors utilized the HTRPM simulator to generate data from seven distinct operational scenarios, which were used to train an LSTM-based fault classification model. The diagnosis module achieved an accuracy of 100\% in identifying the initial fault events. 

Our AutoGraph framework can be incorporated as a modular component within the DRIF architecture, as illustrated in Figure~\ref{DRIF} (b). By integrating DRIF’s native fault diagnosis and checklist generation modules with our AutoGraph, the combined system enables dynamic, real-time decision support for human reliability analysis in digital control room environments. This enabled a fully integrated workflow, linking fault detection, procedural retrieval, and human reliability assessment.

\subsection{Cognitive Simulation via ACT-R for Human Digital Twin Modeling}
As discussed in Section~\ref{Integration with existing DHRA tools}, the AutoGraph framework can be integrated with dynamic human reliability analysis (DHRA) tools such as COGMIF. Beyond this, it can also be combined with cognitive human performance models, including ACT-R, queueing network-model human processor (QN-MHP), and GOMS \cite{boring2016goms}, to construct human digital twins (HDTs) for operator behavior simulation. In this section, we illustrate the integration process using ACT-R as an example.

Building on the capability described in Section~\ref{Automated Execution for Path-Based Parameter Search}, where interface navigation and parameter search are automated through computer control, we propose to simulate the same task execution time using ACT-R. This simulated time can be used not only to support integration with HRA methods like IDHEAS-ECA, as in Section~\ref{Integration with existing DHRA tools}, but also to enforce specific timing constraints in automated interface interactions. In doing so, we achieve spatiotemporal synchronization between human-like cognitive behavior and system-level execution, thereby enabling more realistic simulation of human search and interaction processes in control room environments. 

Importantly, such HDT-based simulations also provide a promising pathway for generating synthetic human reliability data under controlled yet realistic conditions. This can support the development and validation of data-driven HRA methods, particularly in scenarios where real-world behavioral data are limited or difficult to obtain. Moreover, the HDT framework can be further leveraged to evaluate procedural design by simulating the expected execution process and estimating cognitively plausible task durations. This enables the identification of optimal execution times and potential bottlenecks in digital procedures, contributing to procedure refinement and interface optimization.

\section{Conclusions}\label{Conclusion}

In this study, we introduced AutoGraph, a knowledge-graph-based framework designed to model interface interactions and automate procedure execution in digital nuclear control rooms. By integrating interface tracking, semantic–spatial knowledge modeling, and procedural mapping, AutoGraph enables precise representation of control panel elements and supports the automated execution of complex operational tasks. The identification of multi-action steps within procedural paths further enhances the framework’s utility in reducing cognitive workload and minimizing human error.

Through implementation and scenario-based demonstrations, we have shown that AutoGraph is not only effective as a standalone automation system, but also extensible. We explored its integration into the COGMIF framework to support dynamic human reliability assessment, and into the DRIF framework to enhance real-time, fault-driven decision support. These applications underscore the framework’s modularity and potential for broader deployment in safety-critical domains.

Future work will focus on scaling the IE-KG construction process, incorporating adaptive learning mechanisms for interface variation, and conducting large-scale evaluations with diverse user groups. Overall, AutoGraph provides a foundational step toward intelligent, data-driven automation of digital procedure execution in complex socio-technical systems.




\section*{Declarations}

\subsection{Funding}
The research was supported by a grant from the National Natural Science Foundation of China (Grant No. T2192933) and the Foundation of National Key Laboratory of Human Factors Engineering (Grant No. HFNKL2024W07).

\subsection{Conflict of interest}
The authors declare that they have no known competing financial interests.

\subsection{Author contribution} 
Xingyu Xiao: Conceptualization, Methodology, Software, Formal analysis, Data Curation, Visualization, Validation, Writing- Original draft preparation. Peng Chen: Software, Methodology. Jiejuan Tong: Conceptualization, Formal analysis, Supervision, Writing - Review and Editing. Shunshun Liu: Methodology. Hongru Zhao: Supervision, Writing - Review and Editing. Jun Zhao: Supervision, Writing - Review and Editing. Qianqian Jia: Supervision, Writing - Review and Editing. Jingang Liang: Resources, Supervision, Writing - Review and Editing, Project administration, Funding acquisition. Wang Haitao: Supervision, Writing- Reviewing and Editing.

\noindent

\bigskip


\bibliography{sn-bibliography}

\end{document}